\documentclass[9pt,twocolumn,twoside]{opticajnl}
\journal{opticajournal} 

\setboolean{shortarticle}{false}

\usepackage{mathrsfs}
\usepackage{lineno}
\usepackage[ justification=justified]{caption} 
\usepackage{multirow}
\usepackage{titlesec}
\usepackage{tabularx}
\usepackage{physics}
\usepackage{bm}
\usepackage{bbm}
\usepackage{makecell} 
\usepackage[caption=false,subrefformat=parens,labelformat=parens]{subfig}
\usepackage{multibib}

\title{Long-distance discrete-modulated continuous-variable quantum key distribution over 126.56 km fiber with local local oscillator}
\author[1,$\dagger$]{Yan Pan}
\author[2,$\dagger$]{Mingze Wu}
\author[1]{Heng Wang}
\author[2,*]{Junhui Li}
\author[1]{Yun Shao}
\author[1,2]{Jie Yang}
\author[1]{Yaodi Pi}
\author[1]{Ting Ye}
\author[1]{Ao Sun}
\author[3]{Lin Jiang}
\author[3]{Lianshan Yan}
\author[3]{Wei Pan}
\author[1]{Yang Li}
\author[1]{Wei Huang}
\author[2]{Song Yu}
\author[2]{Yichen Zhang}
\author[1,**]{Bingjie Xu}
\affil[1]{National Key Laboratory of Security Communication, Institute of Southwestern Communication, Chengdu 610041, China}
\affil[2]{State Key Laboratory of Information Photonics and Optical Communications, School of Electronic Engineering, Beijing University of Posts and Telecommunications, Beijing 100876, China}
\affil[3]{Center for Information Photonics $\&$ Communications, School of Information Science and Technology, Southwest Jiaotong University, Chengdu, 611756, Sichuan, China}

\affil[$\dagger$]{These authors contributed equally}
\affil{Corresponding authors: $^{*}$lijunhuiphy@bupt.edu.cn, $^{**}$xbjpku@163.com}

\begin{abstract}
	Discrete modulation represents a practical and device-friendly solution for high speed transmission in continuous-variable quantum key distribution (CV-QKD), offering high compatibility with coherent optical communication systems. However, when it comes to transmission distance, discrete-modulated CV-QKD systems fall considerably short compared to their Gaussian-modulated counterparts. Leveraging an advanced security analysis method and probabilistic shaped 16-quadrature amplitude modulation, combined with experimental parameter optimization and an efficient fractional-spacing equalization method based on a variable step-size least mean square (LMS) algorithm, we have experimentally showcased a local local oscillator (LLO) CV-QKD system over 126.56 km span of single-mode fiber with  secret key rate of 171.42 kbps. This system sets a new record of the transmission distance of LLO CV-QKD systems, and provides a promising candidate for the commercialization and practical application of CV-QKD technology.
\end{abstract}

\setboolean{displaycopyright}{false} 

\begin{document}
	
\maketitle
	
\begin{figure*}[!h]
	\centering
	\includegraphics[width=14cm]{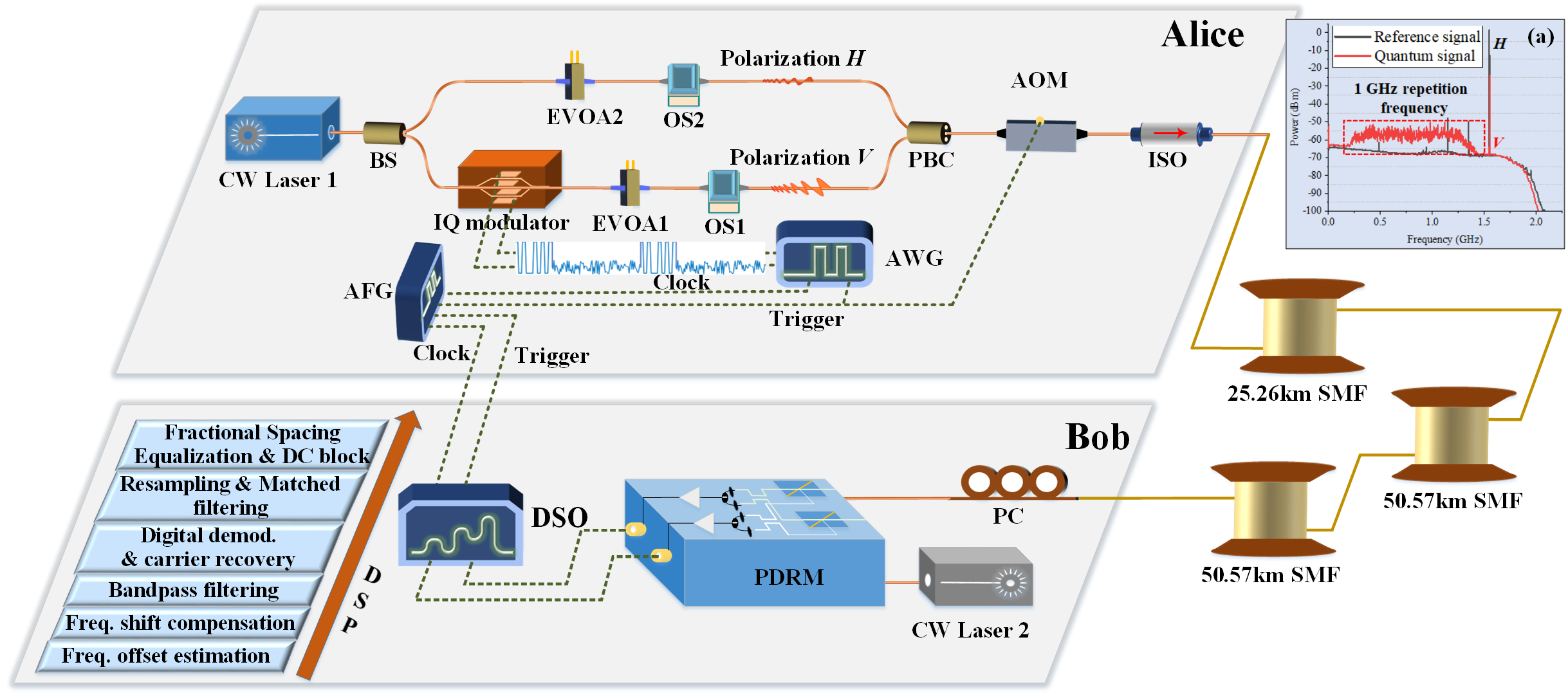}
	\caption{Experimental schematic of the proposed long-distance LLO discrete modulated CV-QKD. CW, continuous-wave; BS, beam splitter; AWG, arbitrary waveform generator; EVOA, electrical variable optical attenuator; OS, optical switch; AFG, arbitrary function generator; PBC, polarization beam combiner; AOM, acousto-optic modulator; ISO, isolator; SMF, single-mode fiber; PC, polarization controller; PDRM, polarization diversity receiver module; DSO, digital storage oscilloscope; DSP, digital signal processing; Fre., frequency. The inset shows the frequency and polarization relationship between quantum signal and reference signal.}
	\label{e1}
\end{figure*}

\section{Introduction}
Quantum key distribution (QKD)\ \cite{1984Quantum} allows unconditional secure distribution of keys between two distant parties theoretically, solving the problem of key transmission in symmetric cryptosystems, and is one of the potential solutions to the threat of quantum computing\ \cite{pirandola2020advances,xu2020secure,portmann2022security}. Utilizing coherent state and coherent detection, continuous-variable (CV) QKD demonstrates advantages of low-cost, more integrable, and high compatibility with coherent optical communication\ \cite{zhang2024continuous}. Consequently, it is regarded as promising candidate for large-scale quantum-safe communication, facilitating seamless integration into existing infrastructure while guaranteeing robust security against potential threats. Particularly, Gaussian modulated CV-QKD is the most popular CV-QKD protocol\ \cite{ralph1999continuous,grosshans2002continuous,weedbrook2004quantum} , which has got comprehensive security analysis\ \cite{leverrier2015composable,leverrier2017security}. 
Meanwhile, numerous theories and approaches have been experimentally validated. Especially in long-distance CV-QKD transmission systems, Gaussian modulation protocols are dominant, either in transmit local oscillator (TLO) or local local oscillator (LLO) schemes \ \cite{jouguet2013experimental,huang2015high,zhang2020long,pi2023,hajomer2024long}.
	
However, limited by practical devices, continuous Gaussian modulation cannot be completely achieved in experiments and needs to be approximated by thousands of constellations\ \cite{jouguet2012analysis}. It faces challenges on implementation complexity, noise sensitivity, and device precision. In comparison with Gaussian modulation, discrete modulation with finite number of coherent states is increasingly recognized for its great potential in long-distance transmission, having advantages on enhanced device compatibility, simpler noise management, etc.
In recent years, discrete-modulated CV-QKD protocol develops rapidly, whose asymptotic security analysis\ \cite{li2018user,leverrier2019,lin2019,denys2021explicit} and composable security analysis\ \cite{lupo2022quantum,kanitschar2023finite,bauml2024security} have been done. The Semidefinite programming (SDP) methods have significantly narrowed the gap in transmission distance compared with Gaussian-modulated protocols. Besides, related practical security researches\ \cite{lin2020trusted,fan2023quantum,wu2024amplitude,wu2024trusted} and protocol improvements\ \cite{liu2021homodyne,upadhyaya2021dimension,kanitschar2022optimizing} lay foundation for practical applications. 

Based on these theoretical results, various protocols with different modulation formats have been experimentally demonstrated\ \cite{wang2022sub,pan2022experimental,tian2023high,roumestan2024shaped,hajomer2024experimental,bian2024continuous}. In Ref. \ \cite{roumestan2024shaped}, F. Roumestan et al. investigated a discrete modulated CV-QKD system by using probabilistic shaped (PS) 64 quadrature amplitude modulation (QAM) and 256QAM, and achieved secret key rate (SKR) of 24 Mbps over transmission distance of 25km single-mode fiber (SMF). Y. Pan et al. demonstrated a PS LLO discrete-modulated CV-QKD system, achieving transmission distance of 50.59 km SMF with SKR of 9.193 Mbps\ \cite{pan2022experimental}. Y. Tian et al. proposed an optimal method for 16 amplitude phase shift keying (APSK) signal in Ref. \ \cite{tian2023high}, and transmitted over 80 km of telecom SMF with SKR of 2.11 Mbps. Unfortunately, no reported work has demonstrated discrete modulated CV-QKD schemes exceeding 100 km of fiber, which lag significantly behind Gaussian modulation protocols.

In this contribution, theoretically, we employ an advanced security analysis method based on nonlinear SDP and optimized the data post-selection, shows significant potential improvement of long distance performance. Besides, we optimize PS-16QAM signals for tighter SKR bound, and propose fractional spacing equalization method based on variable step-size least mean square (LMS) algorithm for effectively suppression of excess noise in long-distance transmissions. Experimentally, a local local oscillator (LLO) discrete-modulated CV-QKD system with 1GBaud PS-16QAM signal is investigated, then under more general non-Gaussian channel, SKR of 171.42 kbps for transmission over 126.56 km of fiber is achieved. To the best of our knowledge, this is the first CV-QKD implementation to realize such long-distance key distribution with discrete modulated protocol, and it also represents the longest LLO CV-QKD system to date. The LLO discrete-modulated scheme explored in this work offers valuable technical support for CV-QKD in metropolitan quantum-safe communication.

\section{Results}

\textbf{System description.} Figure \ref{e1} illustrates experimental setup of the proposed CV-QKD system. At Alice's side, continuous-wave (CW) laser operating at 1550.22 nm with linewidth $\textless$100 Hz serves as the carrier. The laser output is split into two paths using a beam splitter (BS). In one path, the light is launched into an In phase/ Quadrature (IQ) modulator to carry PS-16QAM signal, which is generated by a two channel arbitrary waveform generator (AWG) operating at 10 GSa/s with resolution of 10-bit and symbol rate of 1 Gbaud. Here, discrete Maxwell-Boltzmann distribution (more details can be found in "Methods") and root-raised cosine (RRC) filter with roll-off factor of 0.3 are adopted for waveform shaping. Quadrature phase shift keying (QPSK) training symbols, 16 times the power of quantum signals, are interleaved in the time domain for channel training. An electrical variable optical attenuator (EVOA1) adjusts the modulation variance ($V_A$), and an optical switch (OS1) chooses the quantum link for parameter calibration, together produces discrete modulated coherent state. The reference path is consist of EVOA2 and OS2. Then, the quantum signal and the reference is combined by a  polarization beam combiner (PBC). In this case, the discrete modulated coherent state and reference signals are multiplexed in both polarization and frequency, as depicted in the inset of Fig. \ref{e1}. The multiplexed signal passes through an acoustic-optic modulator (AOM) for real-time shot noise calibration, and an optical isolator (ISO) prevents reflected light interference.
	
The transmission link consists of 126.56 km of commercial ultralow-loss single-mode fiber (SMF) with total loss of 20.50 dB. At Bob's side, an independent CW laser whose linewidth $\textless$100 Hz  serves as the local oscillator (LO), offsets by approximately 1.5 GHz from Alice's laser. The transmitted signal and LO are coherently detected using a polarization diversity receiver module (PDRM), where polarization alignment is achieved with a polarization controller (PC). The PDRM consists of two polarization beam splitters (PBS), two polarization-maintaining optical couplers, and two balanced photodetectors (BPD) with a bandwidth of 1.6 GHz, responsivity of 0.95 A/W, and gain of $3.0\times10^{4}$ V/A. Finally, electrical signals are digitized by a digital storage oscilloscope (DSO) at 8 GSa/s with 10-bit resolution. Offline digital signal processing (DSP) is then performed for raw data recovery and analysis. Here, clock signal for the AWG and DSO is provided by 10 MHz sine wave generated by an arbitrary function generator (AFG). Simultaneously, the AFG outputs a $50\%$ duty cycle pulse signal to trigger the AWG, AOM, and DSO, enabling the synchronized generation, modulation, and acquisition of signals.

\begin{figure}[t]
	\centering
	\includegraphics[width=8.5cm]{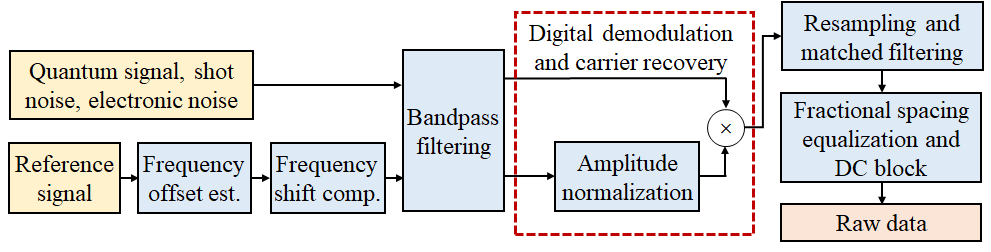}
	\caption{\label{e2}
		The Block diagram of the DSP algorithms.
	}
\end{figure}

\begin{figure}[t]
	\centering
	\includegraphics[width=8.5cm]{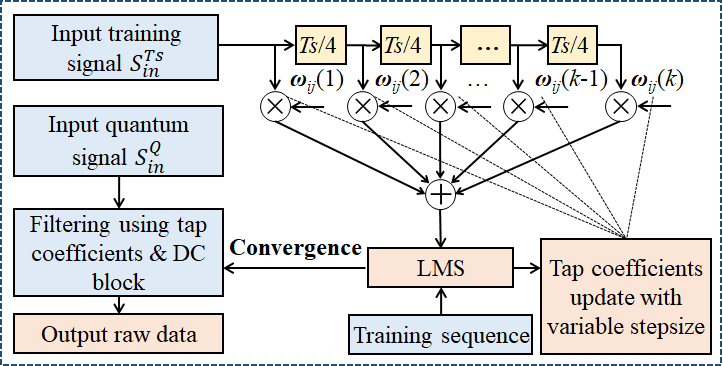}
	\caption{\label{e3}
		Data processing flowchart for fractional spacing equalization. LMS, least lean squares; DC, direct current.
	}
\end{figure}

\begin{figure}[t]
	\centering
	\includegraphics[width=7.5cm]{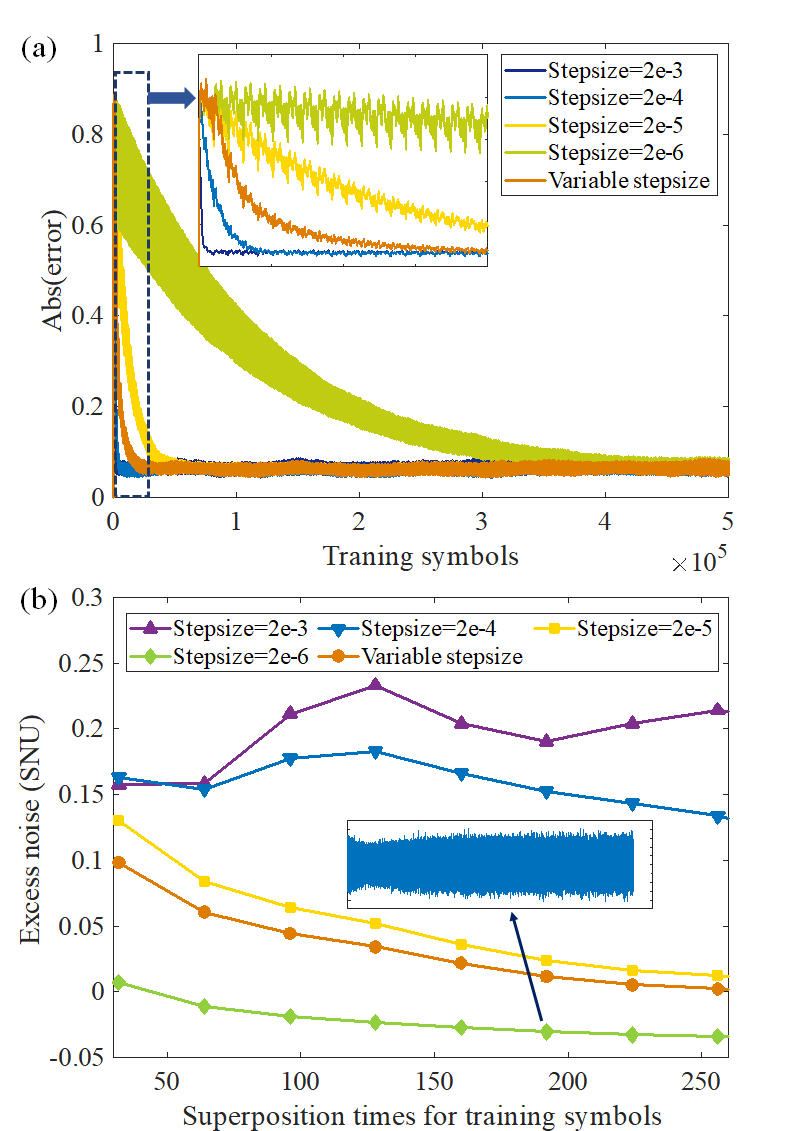}
	\caption{\label{e4}
	Optimization of step-size and superposition times. (a) Convergence of error under different step-size; (b) Excess noise vs. superposition times for training symbols under different step-size. The inset shows the amplitude of the $x$-component of the quantum signal after equalization in the case of step-size set to $2\times 10^{-6}$ and the superposition times set to 192.
	}
\end{figure}

\noindent
\textbf{Excess noise suppression and high precise raw data recovery.} 
For discrete-modulated quantum states, high transmission performance demands not only a sufficiently low statistical variance of excess noise but, more importantly, minimal deviation of quantum signal from the ideal channel response. Therefore, high precise raw data recovery is vital for discrete modulated LLO CV-QKD system, especially when transmission distance is long. However, quantum signals experience various impairments during generation, transmission, and detection, including quantization errors, modulation nonlinearity, link scattering, phase noise, residual polarization disturbances, etc. These challenges make it difficult for the receiver to accurately recover the raw data. In our system, a simple PS-16QAM signal is utilized, allowing quantization error and modulation nonlinearity to be effectively reduced. In addition, as shown in Fig.\ \ref{e2}, a series of efficient DSP algorithms are employed to compensate effects such as link scattering, phase noise, and residual polarization disturbance. The DSP algorithms mainly include frequency offset estimation, frequency shift compensation, bandpass filtering, digital demodulation and carrier recovery, resampling and matched filtering, fractional spacing equalization, and DC block. Here, frequency offset estimation is conducted to find the actual frequency of reference signal. Then, digital frequency shift of 750 MHz is performed for the reference signal to remove center frequency difference between the quantum and reference signal. To suppress out-band noise, frequency-domain ideal filters are used for both signals, with bandwidth of 1.3 GHz and 200 kHz, respectively. In this case, the reference signal and the quantum signal have the same center frequency, and their phase noise is theoretically identical. Therefore, straightforward digital demodulation and carrier recovery method is employed for the quantum signal, as shown in Fig.\ \ref{e2}. This approach not only simplifies the down-conversion process of the quantum signal, but also relaxes accuracy requirements for estimating the reference signal. After that, the demodulated signal is resampled to 4 times oversampling and matched filtered using a root-raised cosine (RRC) filter with roll-off factor of 0.3.

After the above data processing, most signal impairments have been effectively eliminated, resulting in a substantial improvement on the cross-correlation between the processed quantum signal and the transmitter's raw data. However, residual effects such as link-induced nonlinear scattering, phase deviations, $q/p$ component imbalance, and DC bias remain, which significantly contribute to system noise. Here, as shown in Fig.\ \ref{e3}, a fractional spacing equalization using a variable step-size LMS algorithm is proposed to suppress these residual effects. Meanwhile, this algorithm down-samples the quantum signal from 4 times oversampling to one sample per symbol to avoid sampling timing error. Notably, to improve converging speed and accuracy of the LMS algorithm, training sequence enhancement based on the time-domain superposition algorithm is used, and about 200 times superposition in time-domain is applied\ \cite{pan2024high}. Considering intensity at the transmitter was 16 times that of the quantum signal, the final signal-to-noise ratio (SNR) of the training sequence was about 3200 times higher than that of the quantum signal. Here, the intensity and superposition times of training sequence should be optimized according to actual conditions of the transmission link. Additionally, a variable step-size method based on modified Sigmoid function is used, and the convergence of error under different step-size is shown in Fig.\ \ref{e4}(a). To demonstrate the effectiveness of the proposed algorithm, in Fig.\ \ref{e4}(b), the relationship between excess noise and the training sequence superposition times under different step-size is provided. As can be seen from Fig.\ \ref{e4}, larger step-size leads to faster convergence, but higher excess noise due to limited compensation accuracy. In contrast, a smaller step-size improves compensation accuracy at the cost of slower convergence, which may distort the quantum signal and lead to inaccurate excess noise estimation as shown in the inset of Fig.\ \ref{e4}(b). Especially in long-distance CV-QKD systems, where the transmission link may change rapidly, a smaller step-size has difficulty in adaptively equalizing the signal. As shown in Fig.\ \ref{e4}(b), the time-domain superposition algorithm significantly improves excess noise suppression performance with the increasement of superposition times. Therefore, by combining variable step-size equalization with time-domain training sequence superposition, accurate recovery of raw data in long-distance CV-QKD systems can be achieved.

\noindent	
\textbf{Parameter estimation.}
Performance of the PS-16QAM modulated CV-QKD system is shown as follows. In order to evaluate performance of our system, two situations are considered: SKR under Gaussian channel assumption, and that under general condition. Being different from Gaussian modulated CV-QKD, Gaussian attack are not optimal in discrete-modulated CV-QKD protocols. Strictly speaking, we cannot assume the channel to be Gaussian, as this would incorrectly limit the ability of eavesdroppers. It is a general practice to estimate the statistics in SDP directly and calculate SKR accordingly. However, these statistics are not intuitive enough to be compared with Gaussian modulation systems. Besides, in laboratory systems, there is no real eavesdropper. We can consider the channel as an additive Gaussian white noise channel to estimate channel transmittance and excess noise, and then use these two parameters to compare performance and stability of different systems roughly. Hence, both the two performance evaluation methods are used in our work.

Firstly, we use traditional parameter estimation method to estimate channel transmittance $T$ and excess noise $\xi$. We model the quantum channel as an additive Gaussian white noise channel, for heterodyne detection scheme
\begin{equation}
	y={\sqrt{0.5\eta_{d}T}}x+\delta,
\end{equation}
where $x$ and $y$ represent input and output of the channel, $\delta$ is the Gaussian noise with variance $T\eta_d\xi/2+1+\nu_{el}$, $T$ is channel transmittance, $\eta_d$ is detection efficiency, and $\nu_{el}$ is variance of detector electrical noise, respectively. In this case, channel transmittance $T$ and excess noise $\xi$ can be estimated as
\begin{equation}
	T={\frac{\left({\frac{\sum_{i=1}^{m}x_{i}y_{i}}{n}}\right)^{2}}{0.5\eta_{d}}},
\end{equation}
\begin{equation}
	\xi={\frac{V_{B}-0.5\eta_{d}T V_{A}-\nu_{e l}-1}{0.5\eta_{d}T}},
\end{equation}
where $V_{A}$ represents modulation variance of Alice, $V_{B}$ represents variance of Bob's data,  and ${m}$ is symbol number of data used for tests. With channel transmittance and excess noise, we can directly calculate the required statistics for SDP in Gaussian channels as follows\ \cite{lin2020trusted}:
\begin{equation}\label{gc}
	\begin{aligned} 
		&\langle\hat{F}_{Q}\rangle_{x}=\sqrt{2\eta_{d}T}\mathrm{Re}(\alpha_{x}), \\
		&\langle\hat{F}_{P}\rangle_{x}=\sqrt{2\eta_{d}T}\mathrm{Im}(\alpha_{x}),\\
		&\langle\hat{S}_{Q}\rangle_{x}=2\eta_{d}T\mathrm{Re}(\alpha_{x})^{2}+1+\frac{1}{2}\eta_{d}T\xi+\nu_{\mathrm{el}}, \\
		&\langle\hat{S}_{P}\rangle_{x}=2\eta_{d}T\mathrm{Im}(\alpha_{x})^{2}+1+\frac{1}{2}\eta_{d}T\xi+\nu_{\mathrm{el}}, \\
	\end{aligned}
\end{equation}
where $\alpha_{x}$ is the amplitude of each state.

Although the quantum channel can be consider as Gaussian channel in the laboratory systems, in practical situations, the system may be attacked by eavesdroppers, which may be non-Gaussian and results in a lower secret key rate. Therefore, it is necessary to estimate the statistics $\langle \hat{F}_Q \rangle_x$, $\langle \hat{F}_P \rangle_x$, $\langle \hat{S}_Q \rangle_x$, $\langle \hat{S}_P \rangle_x$ required for SDP using experimental data directly, and obtain a more reliable SKR. We first distinguish the measurement data $x^{k}$ and $p^{k}$ for sending each state $\left|\alpha_{k}\right\rangle$, and the statistics $\langle \hat{F}_Q \rangle_x$, $\langle \hat{F}_P \rangle_x$, $\langle \hat{S}_Q \rangle_x$, $\langle \hat{S}_P \rangle_x$ can be calculated as

\begin{equation}\label{gen}
	\begin{aligned} 
		& \langle\hat{F}_{Q}\rangle_{x}^{g}=\frac{1}{C_{k}}\sum\nolimits_{j=1}^{C_{k}}x_{j}^{k}, \\
		& \langle\hat{F}_{P}\rangle_{x}^{g}=\frac{1}{C_{k}}\sum\nolimits_{j=1}^{C_{k}}p_{j}^{k},\\
		& \langle\hat{S}_{Q}\rangle_{x}^{g}=\frac{1}{C_{k}}\sum\nolimits_{j=1}^{C_{k}}\left(x_{j}^{k}\right)^{2}, \\
		&\langle\hat{S}_{P}\rangle_{x}^{g}=\frac{1}{C_{k}}\sum\nolimits_{j=1}^{C_{k}}\left(p_{j}^{k}\right)^{2}, \\
	\end{aligned}
\end{equation}
where $C_{k}$ is the number of rounds sending states $\left|\alpha_{k}\right\rangle$. Based on this, SDP can be calculated to obtain the SKR under general condition.

\begin{figure}[t]
	\centering
	\includegraphics[width=8cm]{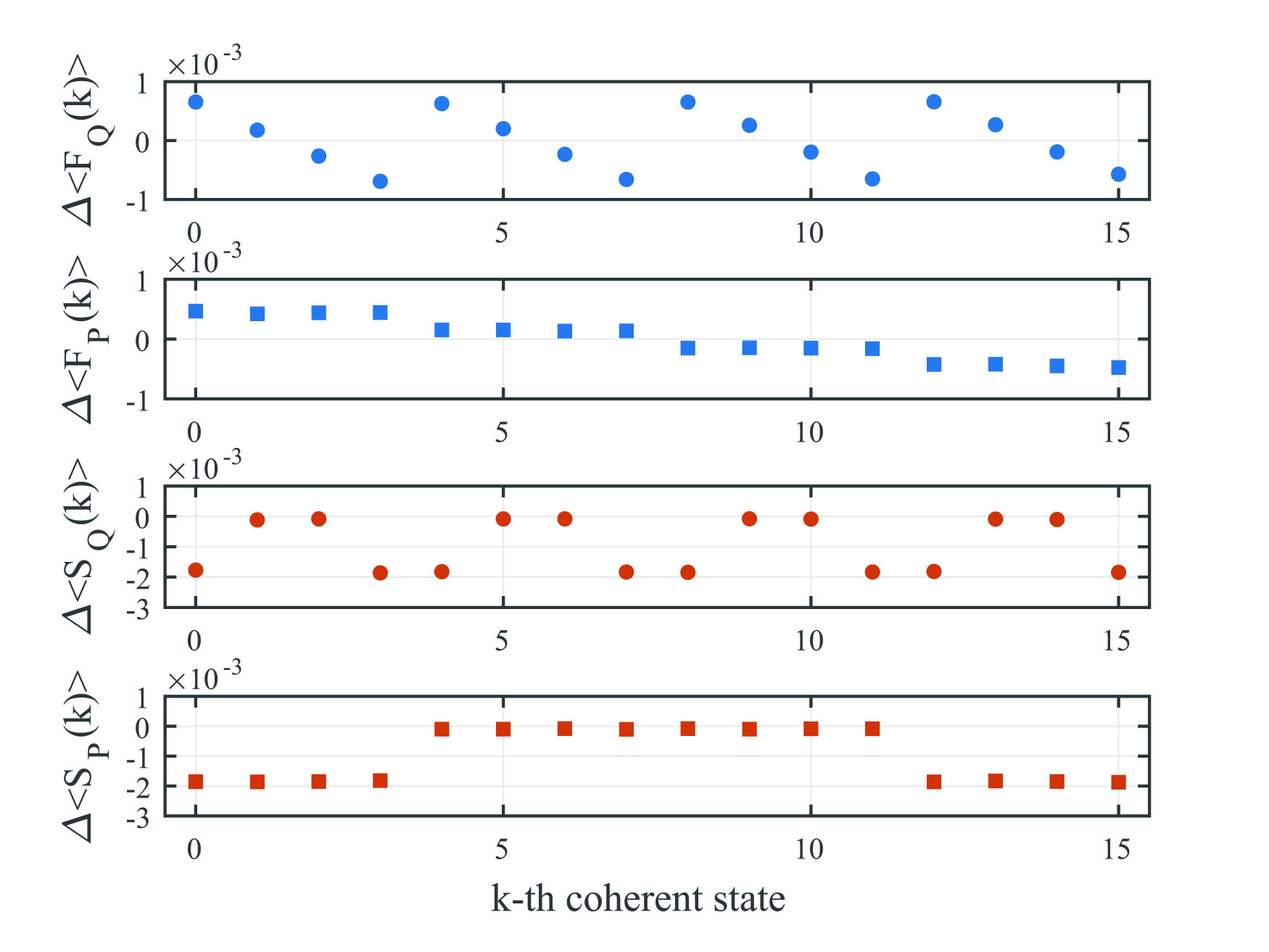}
	\caption{\label{gap}
		Experimental estimated discrepancy  $\Delta\langle \hat{F}_Q \rangle_x$, $\Delta\langle \hat{F}_P \rangle_x$, $\Delta\langle \hat{S}_Q \rangle_x$, $\Delta\langle \hat{S}_P \rangle_x$ of statistics between Gaussian channel and general condition. 
	}
\end{figure}

To get higher SKR, modulation variance $V_A$ and probability distribution parameter $\nu$ in Eq. (\ref{nu}) are optimized. Modulation variance is 2.03 in shot noise unit (SNU) and $\nu=0.2$ for a better SKR. In experiment, power of the quantum signal can be adjusted by high-performance variable optical attenuator (EXFO LTB), and modulation variance can be accurately controlled by meticulously stabilizing the bias of IQ modulator using ultra-high precision bias controller (MBC-IQ-03). With symbol rate $R_s$=1 GBaud, transmission distance $L=126.56$ km, and fiber loss $\alpha=0.162$ dB/km in our experiment, we collect 20 sets of data blocks for parameter estimation, each contains $3.2\times10^8$ symbols. Detection efficiency and electronic noise are calibrated as 0.714 and 0.064 SNU, respectively. Using the parameter estimation method under Guassian channel assumption, excess noise is experimentally estimated to 0.019 SNU, and channel transmittance is estimated to 0.009. The statistics $\langle \hat{F}_Q \rangle_x$, $\langle \hat{F}_P \rangle_x$, $\langle \hat{S}_Q \rangle_x$, $\langle \hat{S}_P \rangle_x$ can be calculated with excess noise and transmittance according to Eq. (\ref{gc}) under Guassian channel assumption. Besides, the general parameter estimation method is also considered in our work according to Eq. (\ref{gen}). The estimated results has some differences with the estimation under Guassian channel assumption. The discrepancy between the two methods is shown in Fig.\ \ref{gap}.


	\begin{table}[ht]
	\centering
	\caption{\textbf{Experimental parameters.}}
	\resizebox{0.98\hsize}{!}{
		\begin{tabular}{|c|c|c|}
			\hline
			\textbf{Parameter}                    & \textbf{Symbol}     & \textbf{Value}                                     \\ \hline
			\textit{\textbf{\textbf{Modulation variance}}} & $V_{A}$         & $2.03$                                           \\ \hline
			\textit{\textbf{Probability distribution parameter}} & $\nu$         & $0.2$                                           \\ \hline
			\textit{\textbf{Fiber loss}} & $\alpha$            & $0.162$ (dB/km)                                           \\ \hline
			\textit{\textbf{Signal-to-noise ratio}} & $SNR$            & 0.0061 (dB)                                          \\ \hline
			\textit{\textbf{Channel transmittance}}                   & $T$                 & $0.009$ \\ \hline
			\textit{\textbf{Excess noise}}                   & $\xi$                 & $0.019$ \\ \hline
			\textit{\textbf{Electronic noise}}                   & $\nu_{\mathrm{el}}$                 & $0.064$ \\ \hline
			\textit{\textbf{Detection efficiency}}                   & $\eta_d$                 & $0.714$ \\
			\hline
			\textit{\textbf{Post-selection parameter}}                   & $\Delta_0$                 & $0.6$ (NU) \\ \hline
			\textit{\textbf{Reconciliation efficiency}}                   & $\beta$                 & $0.95$ \\ \hline
			\textit{\textbf{Frame error rate}}                   & $FER$                 & $0.15$ \\ \hline
			\textit{\textbf{Repetitive frequency}}                   & $R_{s}$                 & 1 (GBaud) \\ \hline	
			\textit{\textbf{Ratio used for parameter estimation}}                   & $a$                 & 10\% \\ \hline	
			\textit{\textbf{Ratio used for training sequence}}                   & $b$                 & 25\% \\ \hline

		\end{tabular}
	}
	\label{tab1}
\end{table}

\begin{figure}[t]
	\centering
	\includegraphics[width=7.5cm]{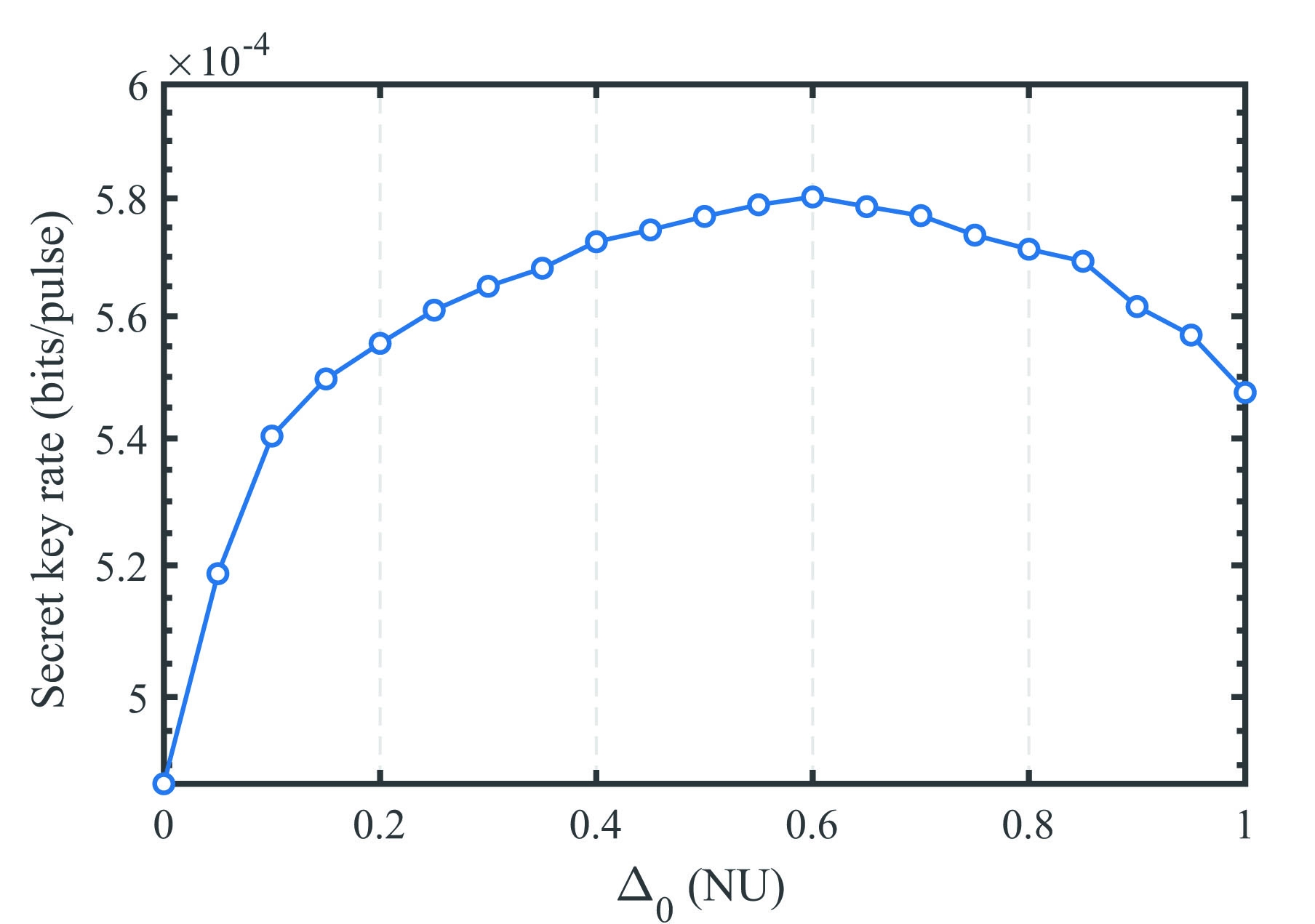}
	\caption{\label{delta}
		Optimization of post-selection parameter $\Delta_{0}$ under our experimental parameters.
	}
\end{figure}

\noindent	
\textbf{Post-selection.} In order to optimize performance of our system, we improve the post-selection process. The motivation of post-selection is to discard data from areas in phase space with high bit-error rates in order to improve SKR, and reduce the computational burden in error correction\ \cite{kanitschar2022optimizing}. One can determine post selection area according to bit error probability directly to the contour lines of the bit-error probability, but it is usually irregular. Balancing effectiveness and complexity, post-selection chooses to discard data near the coordinate axis and corresponding region operators are given in section "Methods". We optimize post-processing parameter $\Delta_{0}$ according to SKR of Gaussian channel as shown in Fig.\ \ref{delta}, where $\Delta:=\sqrt{\eta\eta_{d}}\Delta_{0}$. Results show that the optimal post-selection parameter is around 0.035 NU. It should be noted that the unit used for post-selection parameters is natural units (NU), which follows definition in the security analysis\ \cite{kanitschar2023finite} and differs from the commonly used units for shot noise unit (SNU) in experiments by a factor of $\sqrt{2}$.

Because we don't perform practical error correction and privacy amplification, we set reconciliation efficiency $\beta=0.95$, and frame error rate $FER=0.15$. $10\%$ of raw keys are used for parameter estimation, and $25\%$ of raw keys are used for training sequence. All experimental parameters are summarized in Table\ \ref{tab1}. 

\begin{figure}[t]
	\centering
	\includegraphics[width=7.5cm]{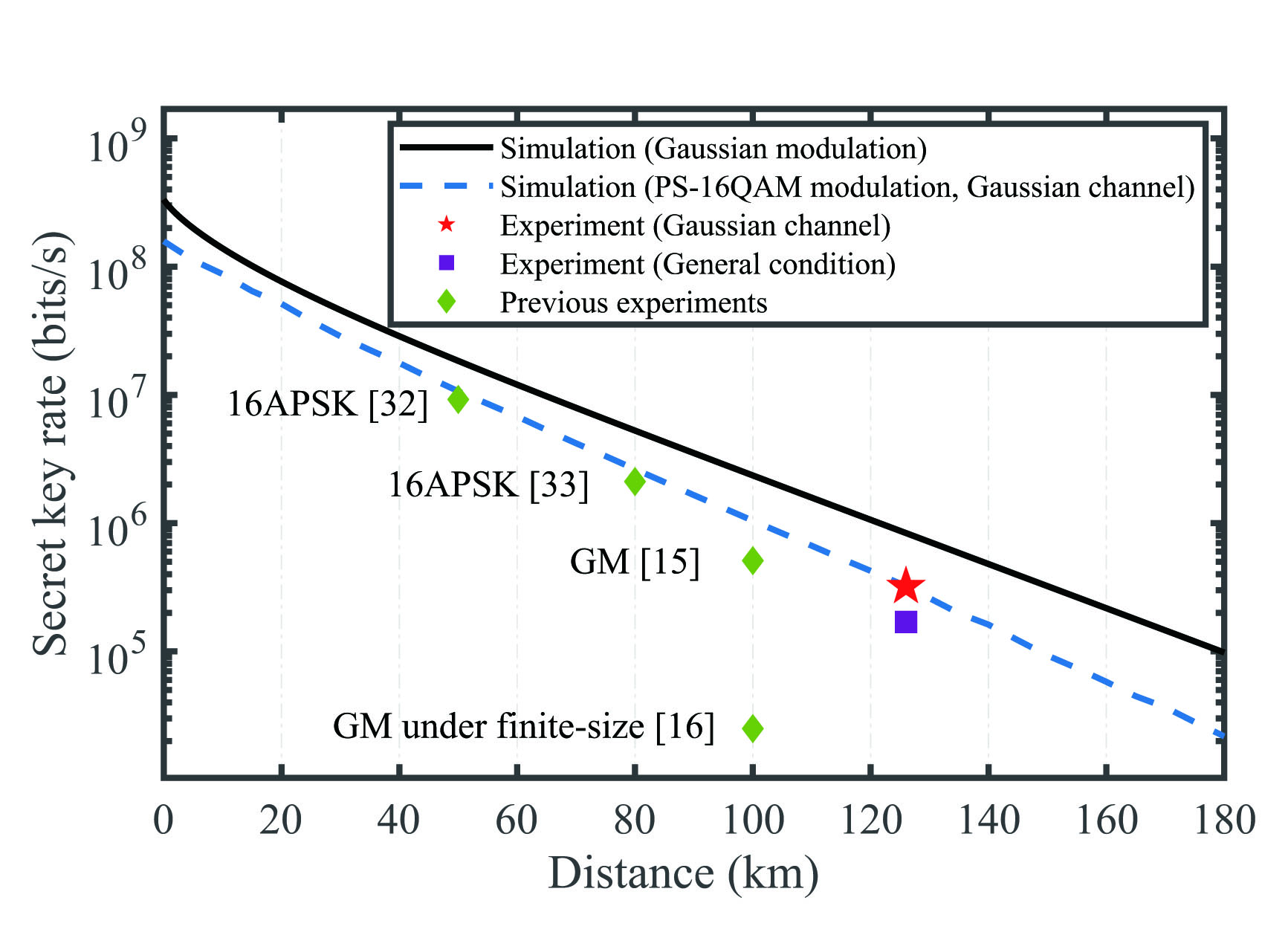}
	\caption{\label{keyrate}
		Secret key rate versus transmission distance with experimental parameters. The black line is the simulation of Guassian modulated protocol using experimental parameters, the blue line is the simulation of PS-16QAM modulated protocol using experimental parameters in Guassian channel, the red star is experimental result under Guassian channel assumption, the purple square is experimental result in general estimation, and the green diamonds are previous representative experimental results\ \cite{pi2023,hajomer2024long,pan2022experimental,tian2023high}.
	}
\end{figure}

The system secret key rate is shown in Fig. \ \ref{keyrate}. The blue line is simulated SKR using experimental parameters estimated under Guassian channel assumption. At the distance of 125.56km, experimental SKR under Guassian channel assumption and general SKR are 322.21 kbps and 171.42 kbps, respectively. Due to the influence of statistical fluctuations, using the general parameter estimation methods resulted in lower SKR. Meanwhile, previous representative experimental results are quoted for reference\ \cite{pi2023,hajomer2024long,pan2022experimental,tian2023high}. It can be seen that the transmission distance of our system exceeds that of all previous LLO CV-QKD systems, and also has certain advantages in terms of SKR. This record breaking result validates the usability of the discrete-modulated CV-QKD protocol in long-distance systems. We can use PS-16QAM modulation to achieve the system performance that is close to, or even surpasses, Gaussian modulation.		

	%

\section{Conclusion}
In summary, we demonstrate a long-distance LLO CV-QKD system, distributing a 1 GBaud PS-16QAM signal. PS-16QAM offers simple implementation and good noise suppression, therefore is suitable for long-distance CV-QKD. Experimental findings reveal that, after 126.56 km optical fiber, the SKR attains 322.21 kbps under phase-insensitive Gaussian channel scenario, and 171.42 kbps under more general non-Gaussian channel condition. In our knowledge, this is the first report of such extended transmission distance for LLO CV-QKD system. The system presented herein stands out as a promising contender for cost-effective, long-distance quantum-safe communication.

However, there are still some practical issues. In our protocol, Bob discretizes results of coherent detection to obtain individual bits, which will form the original key. The problem is that the precise values of his measurement results cannot be used for information reconciliation. For this type of protocol, traditional Gaussian modulation reconciliation scheme is not applicable, and how to achieve efficient information reconciliation is still an open problem\ \cite{leverrier2023information}. Polar code may be a feasible solution to achieve reconciliationn efficiency of $\beta \leq 0.95$\ \cite{jouguet2014high}. In addition, efficient DSP algorithms are crucial for long-distance transmission, and potential optimization algorithms can further enhance system performance. Meanwhile, real-time implementation of these algorithms is pivotal for the practical application of discrete-modulation CV-QKD systems.

\section{Methods}
\textbf{PS-16QAM modulated CV-QKD protocol.} In the prepare-and-measure version of 16QAM-modulated CV-QKD  protocol, Alice prepares one of the sixteen coherent states $(k=0,1,...,15)$ with probability $p_k$ each round and transmits to Bob through quantum channel, where the coherent states are distributed in a square pattern with equal spacing and satisfy the Maxwell-Boltzmann distribution as
	
\begin{equation}\label{nu}
			p_{\mathrm{k}}=\frac{\exp\left(-\nu\left(\mathrm{Re}\left(\alpha_{k}^{2}\right)+\mathrm{Im}\left(\alpha^{2}_{k}\right)\right)\right)}{\sum_{k=0}^{15} \exp\left(-\nu\left(\mathrm{Re}\left(\alpha_{k}^{2}\right)+\mathrm{Im}\left(\alpha^{2}_{k}\right)\right)\right)},
\end{equation}
where $\nu$ is the distribution parameter. After receiving the state, Bob uses a heterodyne detector and obtains measurement results. After $N$ rounds of the two steps, Alice and Bob perform classical post-processing as follows:
	
$(1)$ Announcement and sifting. Alice and Bob identify a small subset of test rounds $\mathcal{I}_{\text{test}}$ used for parameter estimation, and use the remaining rounds $\mathcal{I}_{\text{test}}$ to generate keys. Following the sifting process, Alice obtains her string $X=(x_{1}, ..., x_{m})$ according to the following rule:
	
\begin{equation}\label{eq:1}
			\forall j\in[m] \ \ x_{j}=k, \ \ if \left|\psi_{f(j)}\right\rangle=\left|\alpha_{k}\right\rangle, k=0,1,...,15.
\end{equation}
where $m$ is the size of the set $\mathcal{I}_{\text{test}}$ and $f$ is a function from $[m]$ to $\mathcal{I}_{\text{test}}$.

$(2)$ Parameter estimation. Alice and Bob perform parameter estimation by revealing all information from the rounds designated by the test set $\mathcal{I}_{\text{test}}$. To conduct this analysis, they process the data by calculating the observable measurement, conditioned on each of the four states sent by Alice. Subsequently, they compute the SKR in accordance with the optimization problem. If their analysis indicates that secret keys cannot be produced, they abort the protocol. Otherwise, they move forward.
	
\begin{figure}[t]
	\centering
	\includegraphics[width=6cm]{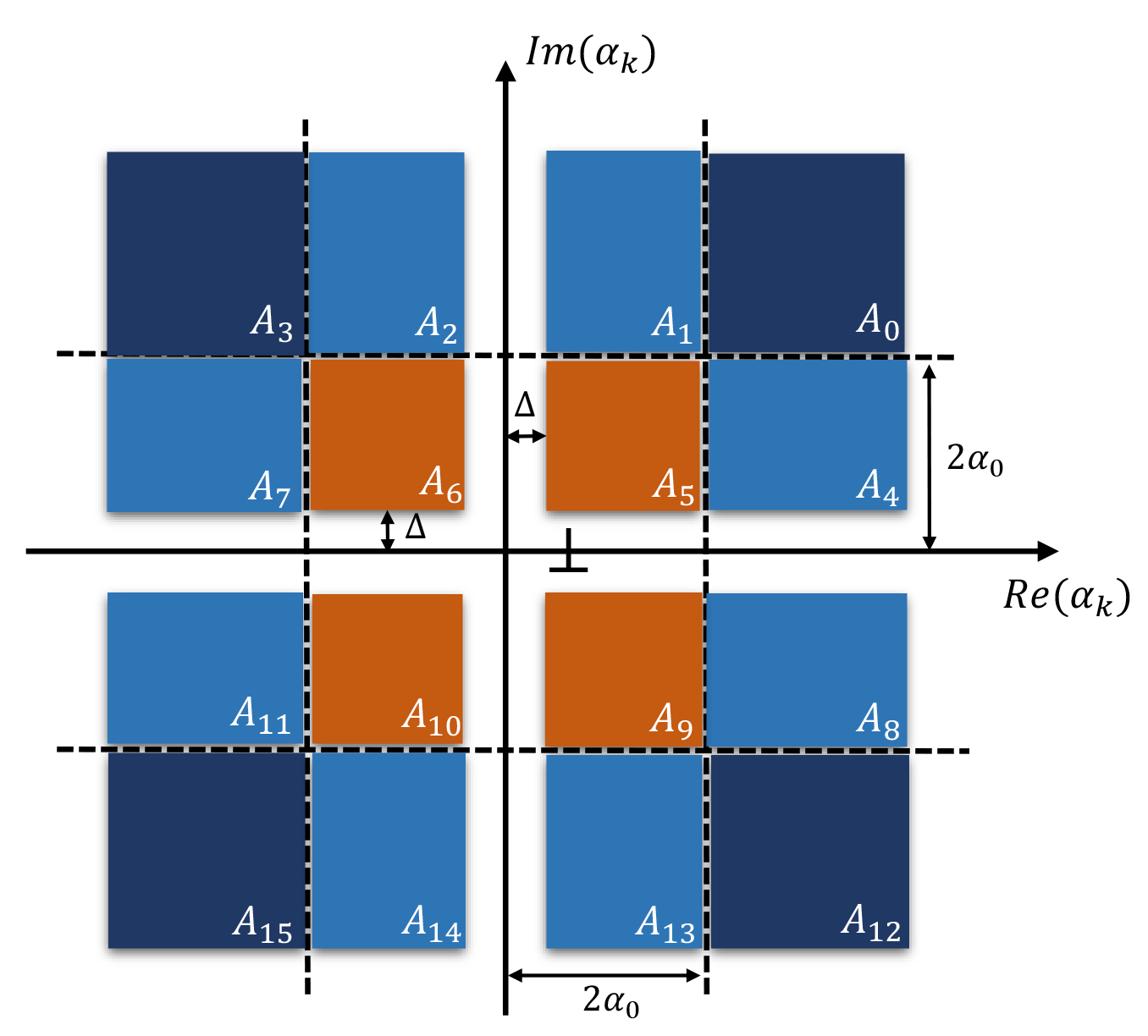}
	\caption{\label{xzt126}
	Bob’s key map process for the measurement results $Y$. Each region $A_{z}$ corresponds to a key map value $z$, $\alpha_0$ is the distance between each adjacent average state at Bob's side. During the post-selection, the measurement result obtained from the range with distance of less than $\Delta$ from the coordinate axis and the range exceeding detection limit are disregarded and instead assigned the symbol $\perp$.
}
\end{figure}

$(3)$ Key map. Bob uses a key map process to derive his raw key string. This map process transforms his measurement result $y_k$ into an element within a specific set $\left\{0, 1, …, 15, \perp\right\}$. Bob obtains his key string $Z =(z_{1}, ..., z_{m})$ according to the rule as shown in Fig.\ \ref{xzt126}
	
$(4)$	Error correction and privacy amplification. Alice and Bob use privacy amplification to diminish Eve’s knowledge of their shared information by eliminating certain portions of their jointly held key.
	
\noindent
\textbf{SKR calculation.} The asymptotic security of our protocol is based on the security analysis using Semidefinite programming (SDP) in\ \cite{lin2020trusted}, and we extend it to PS-16QAM modulation. The SKR formula of discrete modulated CV-QKD protocol is
\begin{equation}\label{eq:1}
	R^{\infty}=\operatorname*{min}_{\rho_{AB}\in s}D\big(\mathcal{G}\big(\rho_{A B}\big)||\mathcal{Z}\big[\mathcal
				{G}\big(\rho_{A B}\big)\big]\big)-p_{pass}\delta_{E C}
\end{equation}
where $p_{pass}$ is the sifting probability and $\delta_{EC}$ is information leakage. The key of the problem lies in the term $\operatorname*{min}_{\rho_{AB}\in s}D\big(\mathcal{G}\big(\rho_{AB}\big)||\mathcal{Z}\big[\mathcal{G}\big(\rho_{A B}\big)\big]\big)$, where $\mathcal{G}$ is a CPTP map that outlines several classical post-processing procedures of the protocol, $\mathcal{Z}$ is a pinching quantum channel that is used to access results of the key map, $D(\rho||\sigma)$  is the quantum relative entropy, and $S$ is the set of density matrices satisfying experimental constraints. For reverse reconciliation, ${\cal G}(\sigma)=K\sigma K^{\dagger}$, where $K=\sum_{z=0}^{15}|z\rangle_{R}\otimes I_{A}\otimes(\sqrt{R_{z}})_{B}$ and $R_{z}$ are region operators. Considering the trusted detection noise, the region operators can be expressed as\ \cite{lin2020trusted}
\begin{equation}
	R_{z}=\textstyle\int_{\zeta\in A_{j}}G_{\zeta}d^{2}\zeta,
\end{equation}
where $A_{j}$ are the regions of post-selection, and $G_{y}$ is the POVM element corresponding to the noisy heterodyne detector
\begin{equation}
	G_{\zeta}=\frac{1}{\eta_{d}\pi}\hat{D}\left(\frac{\zeta}{\sqrt{\eta_{d}}}\right)\rho_{\mathrm{th}}\left(\frac{1-\eta_{d}+\nu_{\mathrm{el}}}{\eta_{d}}\right)\hat{D}^{\dag}\left(\frac{\zeta}{\sqrt{\eta_{d}}}\right),
\end{equation}
where $\eta_d$ is detection efficiency, $\nu_{el}$ is electrical noise, $\hat{D}\left(\cdot\right)$ is displacement operator and $\rho_{\mathrm{th}}\left(\cdot\right)$ is thermal state. For 16QAM modulation, we consider cartesian coordinates, and the region operators become
\begin{equation}
	R_{z}=\int_{\Delta_{xlow}}^{\Delta_{xup}}\int_{\Delta_{ylow}}^{\Delta_{yup}}G_{x+iy} dydx,
\end{equation}
where the integral ranges are
\begin{equation}
	\begin{aligned}
		\Delta_{xlow}=&\left\{2\alpha_{0},\Delta,-2\alpha_{0},-\infty,2\alpha_{0},\Delta,-2\alpha_{0},-\infty, \right.\\
		&\left.2\alpha_{0},\Delta,-2\alpha_{0},-\infty,2\alpha_{0},\Delta,-2\alpha_{0},-\infty \right\}, \\
		\Delta_{xup}=&\left\{\infty,2\alpha_{0},-\Delta,-2\alpha_{0},\infty,2\alpha_{0},-\Delta,-2\alpha_{0},\right.\\
		&\left.\infty,2\alpha_{0},-\Delta,-2\alpha_{0},\infty,2\alpha_{0},-\Delta,-2\alpha_{0} \right\},\\
		\Delta_{ylow}=&\left\{2\alpha_{0},2\alpha_{0},2\alpha_{0},2\alpha_{0},\Delta,\Delta,\Delta,\Delta,\right.\\
		&\left.-2\alpha_{0},-2\alpha_{0},-2\alpha_{0},-2\alpha_{0},-\infty,-\infty,-\infty,-\infty \right\},\\
		\Delta_{yup}=&\left\{\infty,\infty,\infty,\infty,2\alpha_{0},2\alpha_{0},2\alpha_{0},2\alpha_{0},\right.\\
		&\left.-\Delta,-\Delta,-\Delta,-\Delta,-2\alpha_{0},-2\alpha_{0},-2\alpha_{0},-2\alpha_{0}\right\}.
	\end{aligned}
\end{equation}

We need to calculate the region operators $R_{z}$ in photon-number basis. It has a similar approach to the calculations in Ref.\ \cite{kanitschar2022optimizing}, but due to differences in upper and lower bounds, the analytical formulas differ.
We first consider the case $n\leq m$. 
\begin{equation}
	\begin{aligned}
		R=\sum_{n,m}|n\rangle\langle m|C_{n,m}\int_{\Delta_{x l o w}}^{\Delta_{x u p}}\int_{\Delta_{y l o w}}^{\Delta_{y u p}}e^{-a(x^{2}+y^{2})} \\
		(x-i y)^{m-n}L_{n}^{m-n}\big(-\frac{x^{2}+y^{2}}{b}\big)d y d x,
	\end{aligned}
\end{equation}
where $C_{n,m}:=\{1/\pi\,\eta_{d}[(m-n)/2]+1\}\sqrt{n!/m!}(\overline{{n}}_{d}^{n}/(1+\overline{{n}}_{d})^{m+1})$, $a:=1/\eta_{d}(1+\overline{{{n}}}_{d})$, $b:=\eta_{d}\bar{n}_{d}(1\ +\bar{n}_{d})$, $\bar{n}_{d}:=(1-\eta_{d}+\nu_{\mathrm{el}})/\eta_{d}$ and $L_{n}^{a}(x)$ is the generalized Laguerre polynomial of degree $k$ and with parameter $\alpha$.
		
The optimization problem can be expressed as
\begin{equation}\label{opt}
	\begin{aligned} 
		&\text{minimize}\ \ D\big(\mathcal{G}(\rho_{AB}) || \mathcal{Z}[\mathcal{G}(\rho_{AB})]\big)\\
		&\text{subject to} \\
		&\hspace{2em}
	\begin{cases}
	\vspace{4pt}
	& \Tr[\rho_{AB} (\dyad{x}{x}_A \otimes \hat{F}_Q)] = p_x \langle \hat{F}_Q \rangle_x, \\
	\vspace{4pt}
	& \Tr[\rho_{AB} (\dyad{x}{x}_A \otimes \hat{F}_P)] = p_x \langle \hat{F}_P \rangle_x,\\
	\vspace{4pt}
	& \Tr[\rho_{AB} (\dyad{x}{x}_A \otimes \hat{S}_Q)] = p_x \langle \hat{S}_Q \rangle_x, \\
	\vspace{4pt}
	& \Tr[\rho_{AB} (\dyad{x}{x}_A \otimes \hat{S}_P)] = p_x \langle \hat{S}_P \rangle_x, \\
	\vspace{4pt}
	& \Tr[\rho_{AB}] = 1,\\
	\vspace{4pt}
	& \Tr_B [\rho_{AB}] = \sum_{i,j=0}^{15} \sqrt{p_i p_j} \bra{\alpha_j}\ket{\alpha_i} \dyad{i}{j}_{A}, \\ 
	\vspace{4pt}
	& \rho_{AB} \geq 0, 
	\end{cases}
	\end{aligned}
\end{equation}
where $\hat{F}_Q$, $\hat{F}_P$, $\hat{S}_Q$, $\hat{S}_P$ are Bob's observable operators and $\langle \hat{F}_Q \rangle_x$, $\langle \hat{F}_P \rangle_x$, $\langle \hat{S}_Q \rangle_x$, $\langle \hat{S}_P \rangle_x$ are corresponding expectation values. Experimental data need to be used to calculate the statistics required for SDP according to the numerical calculation method\ \cite{coles2016numerical,winick2018reliable}. Since the practical error correction is not preformed in our experiment, we consider the  error-correction cost $\delta_{\mathrm{EC}}$ at the Shannon limit, which can be determined by the joint probability distribution\ \cite{lin2020trusted}
\begin{equation}
	P(z=j|x=k)=\frac{1}{\pi\left(1+\frac12\eta_{d}T\xi+\nu_{\mathrm{el}}\right)}\exp\left[-\frac{\left|y-\sqrt{\eta_{d}T}\alpha_{k}\right|^{2}}{1+\frac12\eta_{d}T\xi+\nu_{\mathrm{el}}}\right]
\end{equation}
where Alice's data after mapping is $k\in\left\{0,1,...,15\right\}$ and Bob's data after key map is $z\in\left\{0,1,...,15,\perp\right\}$. Furthermore, the error correction leakage term can be calculated as
\begin{equation}
	\delta_{EC}=H(Z)-\beta I(X;Z),
\end{equation}
where $H(Z)$ is the Shannon entropy of the raw key $Z$, $\beta$ is the reconciliation efficiency of the chosen error-correction code, and $I(X;Z)$ is the classical mutual information between $X$ and $Z$.

After calculating the theoretical SKR, system SKR can be obtained with the parameters of practical system:
\begin{equation}\label{eq:1}
	R^{\infty}_{sys}=R_{s}(1-a-b)(1-FER)R^{\infty},
\end{equation}
where $R_{s}$ is repetitive frequency of system, $a$ is the overhead ratio for parameter estimation, $b$ is the ratio for training sequence, and FER is the frame error rate.

\begin{backmatter}
	\bmsection{Acknowledgments} We acknowledge financial support from the National Key Research and Development Program of China (Grant No. 2020YFA0309704), the National Natural Science Foundation of China (Grants No. U24B2013, U22A2089, 62471446, 62301517, 62101516, 62171418, 62201530), the Sichuan Science and Technology Program (Grants No. 2024ZYD0008, 2024JDDQ0008, 2023ZYD0131, 2023JDRC0017, 2022ZDZX0009, 2023NSFSC1387, 2024NSFSC0470, and 2024NSFSC0454), the National Key Laboratory of Security Communication Foundation(Grant No. 6142103042301, 6142103042406), Stability Program of National Key Laboratory of Security Communication(Grant No. WD202413, WD202414), National Natural Science Foundation of China (Grant No. 62001044), the Basic Research Program of China (Grant No. JCKY2021210B059), the Equipment Advance Research Field Foundation (Grant No. 315067206).

	\bmsection{Disclosures} The authors declare no conflicts of interest.
	
	\bmsection{Data Availability Statement}	Data underlying the results presented in this paper are not publicly available at this time but may be obtained from the authors upon reasonable request.
	
\end{backmatter}

	\bibliography{ref}

\end{document}